\documentclass[
showpacs,
floatfix,
aps, 
prl,
twocolumn,
superscriptaddress,
amssymb,
]{revtex4-1}

\usepackage{amsmath}

\usepackage{graphicx}
\usepackage{times}
\usepackage{wasysym}
\usepackage{amssymb}
\usepackage{latexsym}

\usepackage{dcolumn}
\usepackage{amsfonts}
\usepackage{bm}
\usepackage{epsfig}

\def\degree{^{\circ}}

\newcommand{\be}{\begin{equation}}
\newcommand{\ee}{\end{equation}}
\newcommand{\bea}{\begin{eqnarray}}
\newcommand{\eea}{\end{eqnarray}}

\newcommand{\rfig}[1]{Fig.\,\ref{#1}}
\newcommand{\rfigss}[1]{Figs.\,\ref{#1}}

\newcommand{\rref}[1]{Ref.\,\onlinecite{#1}}

\def\ne{n_e}

\def\rxx{R_{xx}}
\def\ryy{R_{yy}}

\def\rh{R_H}

\def\bip{B_\parallel}
\def\easy{\left < 110 \right >}
\def\hard{\left < 1\bar10 \right >}
\def\x{\hat{x}}
\def\y{\hat{y}}

\def\ne{n_e}

\begin{document}
\title{Anomalous nematic-to-stripe phase transition driven by in-plane magnetic fields}
\author{X. Fu$^\S$}
\affiliation{School of Physics and Astronomy, University of Minnesota, Minneapolis, Minnesota 55455, USA}
\author{Q. Shi$^\S$}
\altaffiliation[Present address: ]{Department of Physics, Columbia University, New York, NY, USA}
\affiliation{School of Physics and Astronomy, University of Minnesota, Minneapolis, Minnesota 55455, USA}
\author{M. A. Zudov}
\email[Corresponding author: ]{zudov001@umn.edu}
\affiliation{School of Physics and Astronomy, University of Minnesota, Minneapolis, Minnesota 55455, USA}
\author{G.\,C. Gardner}
\affiliation{Microsoft Quantum Laboratory Purdue, Purdue University, West Lafayette, Indiana 47907, USA}
\affiliation{Birck Nanotechnology Center, Purdue University, West Lafayette, Indiana 47907, USA}
\author{J.\,D. Watson}
\altaffiliation[Present address: ]{Microsoft Station-Q at Delft University of Technology, 2600 GA Delft, The Netherlands}
\affiliation{Birck Nanotechnology Center, Purdue University, West Lafayette, Indiana 47907, USA}
\affiliation{Department of Physics and Astronomy, Purdue University, West Lafayette, Indiana 47907, USA}
\author{M.\,J. Manfra}
\affiliation{Microsoft Quantum Laboratory Purdue, Purdue University, West Lafayette, Indiana 47907, USA}
\affiliation{Birck Nanotechnology Center, Purdue University, West Lafayette, Indiana 47907, USA}
\affiliation{Department of Physics and Astronomy, Purdue University, West Lafayette, Indiana 47907, USA}
\affiliation{School of Electrical and Computer Engineering and School of Materials Engineering, Purdue University, West Lafayette, Indiana 47907, USA}
\author{K. W. Baldwin}
\affiliation{Department of Electrical Engineering, Princeton University, Princeton, New Jersey 08544, USA}
\author{L. N. Pfeiffer}
\affiliation{Department of Electrical Engineering, Princeton University, Princeton, New Jersey 08544, USA}
\author{K. W. West}
\affiliation{Department of Electrical Engineering, Princeton University, Princeton, New Jersey 08544, USA}
\received{\today}

\begin{abstract}
Anomalous nematic states, recently discovered in ultraclean two-dimensional electron gas, emerge from quantum Hall stripe phases upon further cooling. 
These states are hallmarked by a local minimum (maximum) in the hard (easy) longitudinal resistance and by an incipient plateau in the Hall resistance in nearly half-filled Landau levels. 
Here, we demonstrate that a modest in-plane magnetic field, applied either along $\easy$ or $\hard$ crystal axis of GaAs, destroys anomalous nematic states and restores quantum Hall stripe phases aligned along their native $\easy$ direction.
These findings confirm that anomalous nematic states are distinct from other ground states and will assist future theories to identify their origin.

\end{abstract}

\maketitle
Two-dimensional electrons in GaAs quantum wells can support a variety of phases when subjected to quantizing magnetic fields and low temperatures.
At half-integer filling factors $(\nu = i/2, i = {\rm odd})$, these states include composite fermion metals ($i = 1,3$) \cite{jain:1989}, quantum Hall insulators ($i = 5,7$) \cite{willett:1987,pan:1999b}, and quantum Hall stripe (QHS) phases ($i = 9,11,...$) \cite{koulakov:1996,moessner:1996,fogler:1996,note:fradkin,note:hqhs}.
The QHS phases  can be viewed as unidirectional charge density waves composed of strips with alternating integer filling factors, $\nu = (i \pm 1)/2$.
Being characterized by large resistance anisotropies ($\rxx \gg \ryy$) \citep{lilly:1999a,du:1999}, the QHS phases are usually aligned along $\y \equiv \easy$ crystal axis of GaAs, for yet unknown reason \citep{koduvayur:2011,sodemann:2013,liu:2013,pollanen:2015}. 
Recent experiments \cite{fu:2020a} have shown that some QHS phases ($i = 13,15,17$), once formed at $\sim 0.1$ K, can evolve into anomalous nematic states (ANSs) upon further cooling. 

The ANSs are distinguished from the QHS phases by \emph{opposite} dependencies of the $\rxx$ and the $\ryy$ on the detuning from half-filling $|\delta \nu| \equiv |\nu - i/2|$ and on the temperature $T$ \citep{fu:2020a,note:ans}.
In particular, unlike the QHS phases exhibiting a maximum (minimum) in the $\rxx$ ($\ryy$) at $\delta \nu \approx 0$, the ANSs are marked by a minimum (maximum) in the $\rxx$ ($\ryy$) and exhibit much smaller anisotropy ratio $\rxx/\ryy > 1$. 
In addition, the Hall resistance $\rh$ near $\nu = i/2$ develops a plateau-like feature with the value close to $\rh = 2R_{\rm K}/i$, where $R_{\rm K} = h/e^2$ is the von Klitzing constant.
As shown in \rref{fu:2020a}, a small detuning of $|\delta \nu| \approx 0.08$ transforms the ANS into the QHS phase, reflecting a tight competition between these two ground states.
Such sensitivity to $\delta \nu$ is well documented in the lower spin branch of the $N = 1$ Landau level.
Here, within the range of $0 \le \delta \nu \le 0.1$, one finds \citep{deng:2012a} fragile quantum Hall states at $\nu = 5/2$ \citep{willett:1987,pan:1999b} and $\nu = 32/13$ \citep{kumar:2010}, the reentrant integer quantum Hall state at $\nu \approx 2.43$ \citep{lilly:1999a,du:1999,cooper:1999,eisenstein:2002}, and yet another quantum Hall state at $\nu = 12/5$ \citep{xia:2004}.
The $\nu = 5/2$ state can also be altered by an in-plane magnetic field which can transform it into the QHS phase \citep{pan:1999,lilly:1999b}, isotropic liquid \citep{xia:2010}, or make it nematic \cite{liu:2013b}.

In this Letter we report on the response of the ANSs to in-plane components of the magnetic-fields $\bip = B_x$ and $\bip = B_y$.
We find that the immediate effects of $\bip$ are to transform the minimum (maximum) in the $\rxx$ ($\ryy$) at half-filling into a maximum (minimum), to eliminate the plateau in $\rh$, and to restore the ratio $\rxx/\ryy$ to values consistent with the QHS phases.
Remarkably, the ANSs respond to $\bip$ in essentially the same manner when $\bip$ is applied along either the $\x \equiv \hard$ or the $\y \equiv \easy$ direction;
 in both cases the revived QHS phase is aligned along its native $\easy$ crystal axis.
This is in contrast with the effect of $\bip$ on the QHS phases which respond very differently to $B_x$ and $B_y$ whereas persisting to much higher $\bip$ \citep{pan:1999,lilly:1999b,shi:2016c}.
These observations signal that a modest $\bip \approx 0.5$ T is enough to tip a delicate balance between the ANSs and the QHS phases in favor of the latter, a finding which should be taken into account by theories aimed to explain the origin of the ANSs.

Our sample is a 30 nm-wide GaAs quantum well surrounded by Al$_{0.24}$Ga$_{0.76}$As barriers.
The electrons are supplied by Si doping in narrow GaAs doping wells, sandwiched between thin AlAs layers, which are positioned at a setback distance of 80 nm on both sides of the main GaAs well.
After a brief illumination at $T \approx 5$ K, the electrons had a density $\ne \approx 3.0 \times 10^{11}$ cm$^{-2}$ and a mobility $\mu \gtrsim 2 \times 10^7$ cm$^2$V$^{-1}$s$^{-1}$.
Samples were $4\times 4$ -mm squares with eight indium contacts at the corners and the midsides. 
The longitudinal resistances, $\rxx$ and $\ryy$, were measured using a four-terminal, low-frequency (a few hertz) lock-in technique.
The excitation current was sent through the center of the sample, i.e., between mid-side contacts along $\x$ or $\y$ direction.
The in-plane magnetic field $B_x$ or $B_y$ was introduced by tilting the sample about either $\y$ or $\x$ axis, in separate cooldowns.

\begin{figure}[t]
\includegraphics{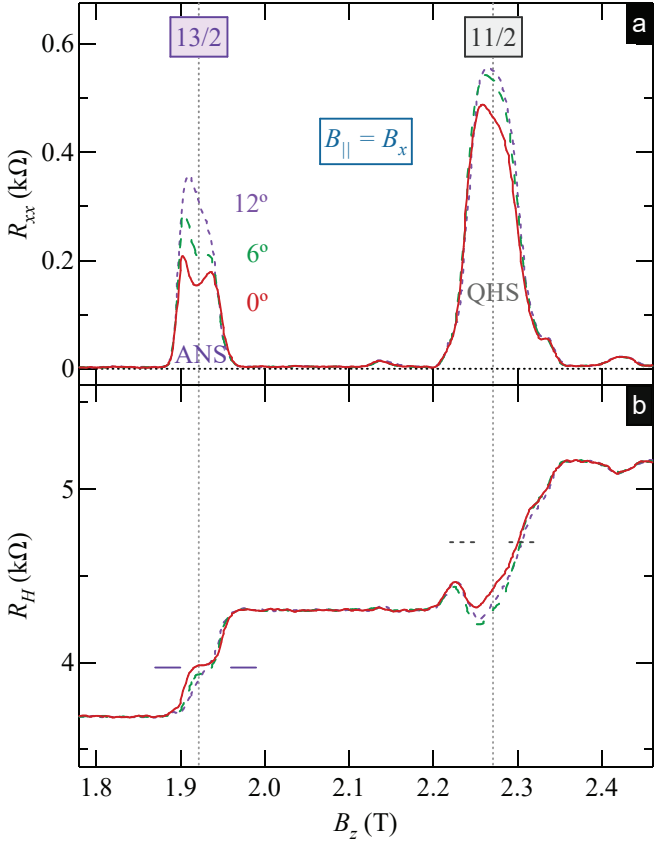}
\vspace{-0.15 in}
\caption{
(a) $\rxx$ and (b) $\rh$ as a function of $B_z$ measured under $\bip = B_y$ at $\theta = 0\degree$ (solid line), $6\degree$ (dashed line), and $12\degree$ (dotted line) at $T \approx 20$ mK. 
Horizontal solid (dashed) lines in (b) are drawn at $\rh = 2R_{\rm K}/13$ ($\rh = 2R_{\rm K}/11$), where $R_{\rm K} = h/e^2$ is the von Klitzing constant.
}
\vspace{-0.15 in}
\label{figx}
\end{figure}
We start with the discussion of the experiments under $\bip = B_x$ (i.e., applied along the $\hard$ crystal axis) and compare its effects on the QHS phase and on the ANS.
In \rfigss{figx}\,(a) [1(b)] we show the longitudinal [Hall] resistance $\rxx$ [$\rh$] as a function of the perpendicular component of the magnetic field $B_z$ at different tilt angles $\theta = 0\degree$ (solid lines), $6\degree$ (dashed lines), and $12\degree$ (dotted lines). 
Consistent with findings of \rref{fu:2020a}, the data at $\bip = 0$ reveal the QHS phase at $\nu = 11/2$ and the ANS at $\nu = 13/2$.
The ANS at $\nu = 13/2$ is evidenced by (i) a minimum in the $\rxx$, with the resistance value much smaller than typical of a QHS phase, and by (ii) an incipient plateau in the $\rh$ with the value close to $\rh = 2R_{\rm K}/13$, as marked by horizontal solid lines.
None of these features are present at $\nu = 11/2$ where the data reflect a conventional QHS phase.

Upon tilting the sample to introduce $\bip = B_x$, the $\rxx$ minimum  at $\nu  = 13/2$ becomes less pronounced at $\theta = 7\degree$ and disappears completely at $\theta = 12\degree$ \citep{note:shift}.
The plateau in the $\rh$ is also destroyed by $B_x$ at this filling factor.
As a result, both the $\rxx$ and the $\rh$ at $\nu = 13/2$ become akin to those at $\nu = 11/2$, hinting at a $\bip$-driven transition from the ANS to the QHS phase.
In contrast, the data at $\nu = 11/2$, representing the QHS phase, exhibit no qualitative changes with increasing $B_x$ in this range of tilt angles, despite higher $B_x$.

\begin{figure}[t]
\includegraphics{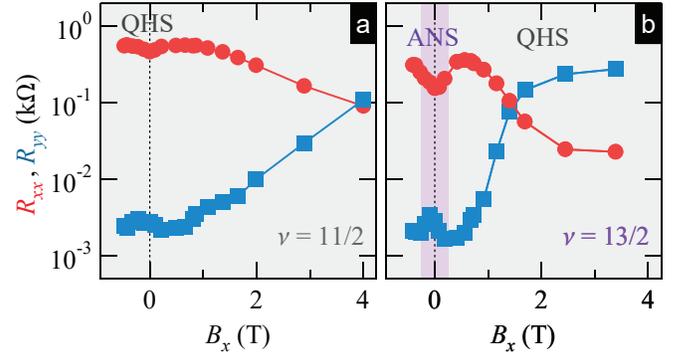}
\vspace{-0.15 in}
\caption{
(a) $\rxx$ (circles), $\ryy$ (squares) vs. $B_x$ at $\nu \approx 11/2$. 
(b) Same as (a) but at $\nu \approx 13/2$.
}
\vspace{-0.15 in}
\label{figx2}
\end{figure}

To further analyze the effects of $\bip = B_x$ on the QHS phase at $\nu = 11/2$ and on the ANS at $\nu = 13/2$ we construct \rfigss{figx2}\,(a) and \rfigss{figx2}\,(b), respectively, which show the $\rxx$ (circles) and the $\ryy$ (squares) as a function of $B_x$ covering a wider range of tilt angles. 
The $\rxx$ ($\ryy$) at $\nu = 11/2$ remains nearly unchanged up to $B_x \approx $ 1 T and then gradually decays (grows) until the anisotropy disappears at $B_x \approx 4$ T, in agreement with earlier experiments \citep{shi:2016c}.
In contrast, the $\rxx$ at $\nu = 13/2$ shows a pronounced maximum at $B_x \approx 0.5$ T, whereas the $\ryy$ shows a deep minimum at the same $B_x$.
As a result, the anisotropy ratio $\rxx/\ryy$ increases 
considerably and becomes consistent with the value exhibited by the QHS phase at $\nu = 11/2$.
Upon further increase of $B_x$, both the $\rxx$ and the $\ryy$ at $\nu = 13/2$ evolve as expected for the QHS phase; here, the anisotropy vanishes at $B_x \approx 1.5$ T, which is lower than the corresponding $B_x$ at $\nu = 11/2$ and the QHS phase realigns along the $\hard$ axis at still higher $B_x$, as anticipated \citep{shi:2016c}.

\begin{figure}[t]
\includegraphics{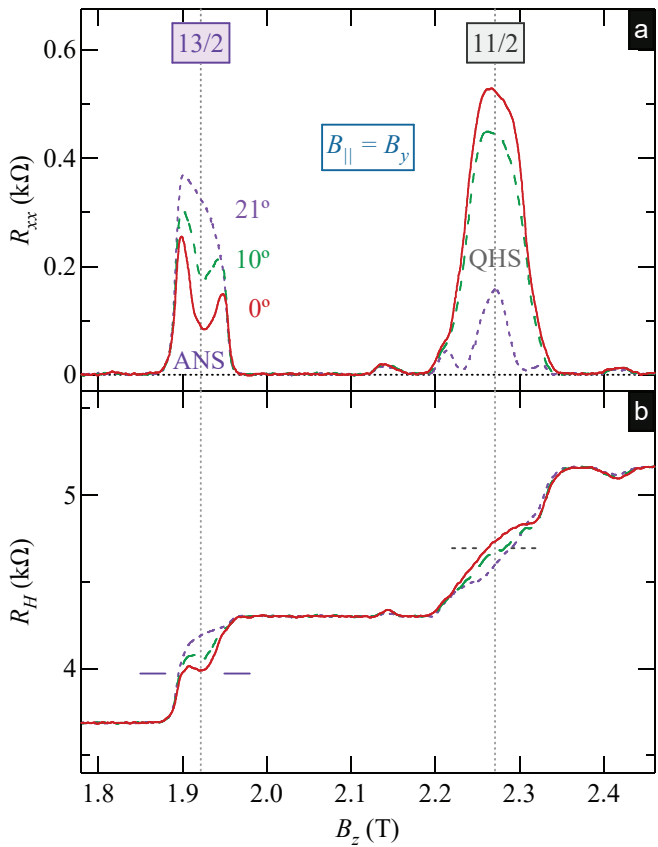}
\vspace{-0.15 in}
\caption{ 
(a) $\rxx$ and (b) $\rh$ as a function of $B_z$ under $\bip = B_y$ at $\theta = 0\degree$ (solid line), $10\degree$ (dashed line), and $21\degree$ (dotted line) at $T \approx 20$ mK. 
Horizontal solid (dashed) lines in (b) are drawn at $\rh = 2R_{\rm K}/13$ ($\rh = 2R_{\rm K}/11$).
}
\vspace{-0.15 in}
\label{figy}
\end{figure}

To acquire further support to the $\bip$-driven ANS-to-QHS phase transition at $\nu = 13/2$ we next present the data under orthogonal orientation of $\bip$ with respect to the anisotropy axis, $\bip = B_y$.
In \rfig{figy}\,(a) we show the $\rxx$ as a function of $B_z$ at $T \approx 20$ mK under $\bip = B_y$ (i.e., applied along $\easy$ crystal axis) at three tilt angles, $\theta = 0\degree$ (solid line), $10\degree$ (dashed line) and $21\degree$ (dotted line).
The Hall resistance data are presented in \rfig{figy}\,(b).
At $\theta = 0\degree$, the data are similar to those obtained in another cooldown, cf. \rfig{figx};
both the $\rxx$ and the $\rh$ exhibit all characteristic features of the QHS phase at $\nu = 11/2$ and of the ANS at $\nu = 13/2$.
It is also evident that in this cooldown the ANS is better developed, as evidenced by a deeper minimum in the $\rxx$.
As discussed in \rref{fu:2020a}, the strength of the ANS sensitively depends on the details of the cool down and illumination protocols.

The evolution of the $\rxx$ in the QHS phase at $\nu = 11/2$ with $B_y$ is consistent with previous studies \citep{shi:2016b,shi:2016c}.
Upon tilting to $\theta = 10\degree$, the $\rxx$ decreases, as the QHS phase starts to reorient perpendicular to $B_y$ \citep{pan:1999,lilly:1999b,shi:2016b,shi:2016c}, and at $\theta = 21 \degree$ the $\rxx$ is reduced much more. 
In contrast to the $\rxx$ at $\nu = 11/2$, the $\rxx$ at $\nu = 13/2$ {\em grows} with the tilt angle and the characteristic ANS minimum quickly disappears. 
Indeed, at $\theta = 21\degree$ the $\rxx$ near $\nu = 13/2$ exhibits a single maximum, as expected of a QHS phase. 
As shown in \rfig{figy}\,(b) the Hall plateau with $\rh \approx 2R_{\rm K}/13$ is also destroyed by $B_y$.
We thus conclude that the effect of $B_y$ on the ANS is essentially the same as that of $B_x$; in either case, the ANS yields to the QHS phase once a modest $\bip$ is introduced.

\begin{figure}[t]
\includegraphics{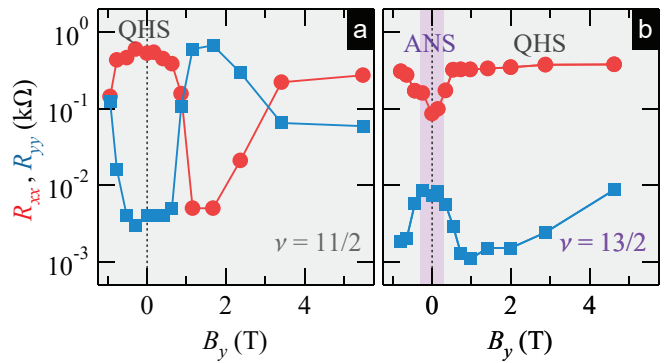}
\vspace{-0.15 in}
\caption{
(a) $\rxx$ (circles), $\ryy$ (squares) vs $B_y$ at $\nu \approx 11/2$. 
(b) Same as (a) but at $\nu \approx 13/2$.
}
\vspace{-0.15 in}
\label{figy2}
\end{figure}

In \rfig{figy2} we summarize the evolutions of both the $\rxx$ (circles) and the $\ryy$ (squares) near (a) $\nu = 11/2$ and (b) near $\nu = 13/2$ over the whole range of $B_y$ studied.
The data at $\nu = 11/2$ reveal two reorientations of the QHS phase; $B_y$ first realigns the QHS phase along the $\hard$ crystal axis (perpendicular to $\bip$) at $B_y \approx 1$ T and then back to along the $\easy$ axis (parallel to $\bip$) at $B_y \approx 3$ T, as previously reported \citep{shi:2016c,shi:2017c}.
At $\nu = 13/2$, however, the data in \rfig{figy2}\,(b) show that the immediate effect of $B_y$ is to dramatically \emph{increase} (\emph{decrease}) the $\rxx$ ($\ryy$) to a value consistent with the QHS phase \citep{note:sbp2}.
As a result of these changes, the resistance anisotropy ratio grows from $\rxx/\ryy \approx 10$ at $B_y = 0$ to $\rxx/\ryy \approx 300$ at $B_y \approx 1$ T.
This fact further supports the $B_y$-driven ANS-to-QHS transition, consistent with our findings under small $\bip = B_x$.

Having concluded that both $B_x$ and $B_y$ transform the ANS to the QHS phase with native orientation, we next comment on possible mechanisms behind this transition.
Let us first assume that the energetics of the ANS phase is not altered by $\bip$. 
According to the calculations of the QHS phases under $\bip$ \citep{jungwirth:1999,stanescu:2000}, $\bip$ serves as an external symmetry-breaking field which either competes with ($\bip = B_y$) or assists ($\bip = B_x$) the native field, responsible for the QHS phase alignment along the $\easy$ direction at $\bip = 0$.
As a result, $B_x$ $(B_y)$ lowers (raises) the energy of the QHS phase with native orientation with respect to its value at $\bip = 0$.
Consistent with this picture,our QHS phase data at $\nu = 11/2$ indeed show that $B_x$ ($B_y$) initially raises (lowers) the anisotropy ratio $\rxx/\ryy$. 
The situation at $\nu = 13/2$, however, is markedly different.
Here, the QHS phase should win over (lose to) the competing ANS as it must become more (less) energetically favorable under modest $B_x$ $(B_y)$.
Although this prediction is consistent with our data under $B_x$, it clearly contradicts our observation of the ANS-to-QHS phase transition under $B_y$.
Furthermore, even though the theory \citep{jungwirth:1999,stanescu:2000} dictates that $B_x$ lowers the energy of the QHS phase with its native orientation, all previous transport studies \citep{lilly:1999b,shi:2016c} have shown that the anisotropy ratio $\rxx/\ryy$ at $\nu = 13/2$ is, in fact, {\em reduced} by $B_x$.
Our data at $\nu = 13/2$, on the other hand, clearly show significant increase in the $\rxx/\ryy$ once $B_x$ is turned on.
Based on the above arguments, we can conclude that $\bip$, regardless of its orientation, raises the energy of the ANS above its value at $\bip = 0$, making it less favorable than the QHS phase at modest $\bip$.

Distinct effects of $B_y$ on the ANS and on the QHS phase are further highlighted by  opposite dependencies on the detuning $|\delta \nu|$ from half-filling.
Indeed, near $\nu = 13/2$, the response of the ANS to $B_y$ is obviously the strongest at $\delta \nu = 0$ resulting in the disappearance of the $\rxx$ minimum and the restoration of a single maximum.
This is in contrast to the QHS phase near $\nu = 11/2$, at which two deep $\rxx$ minima emerge \emph{away} from half-filling at $\theta = 21 \degree$.
These minima appear because lower $B_y$ is required to reorient the QHS phase perpendicular to $\bip$ away from half-filling than at $\delta \nu = 0$ \citep{shi:2016b}.

It is interesting to note that the local minimum in the hard resistance and the  maximum in the easy resistance, as found in the ANS at $\delta \nu = 0$ without an in-plane magnetic field, can also be realized under $\bip = B_y$ when the QHS phase is about to complete its reorientation perpendicular to $\bip$, see, e.g., Fig.\,1(e) in \rref{shi:2016b}.
Such dependencies on $\delta \nu$ can be attributed to a possible decrease in the native symmetry breaking field as one moves away from half-filling \citep{shi:2016b}.
However, if one were to treat the ANS as the QHS phase, the opposite conclusion emerges. 
Indeed, both the reduced anisotropy at $\bip = 0$ and the stronger effects of $\bip$at $\delta \nu = 0$ would suggest that the native field is the weakest at half-filling.
It seems unlikely that the $\delta \nu$ dependencies of the native field be so drastically different at $\nu = 13/2$ and at $\nu = 11/2$ \citep{note:corr}.

To summarize, we have observed a transition from the anomalous nematic state to the quantum Hall stripe phase upon application of the a modest in-plane magnetic field, highlighting tight competition between these two ground states.
The transition occurs both when $\bip$ is aligned along the $\easy$ and along the $\hard$ crystal axis of GaAs and the resultant quantum Hall stripe phase is aligned along native the $\easy$ direction.
Our analysis suggests that $\bip$ likely raises the energy of the ANS compared to its value at $\bip = 0$.
These findings further distinguish anomalous nematic states from other ground states in half-filled Landau levels.

\begin{acknowledgments}
We thank G. Jones, S. Hannas, T. Murphy, J. Park, A. Suslov, and A. Bangura for technical support.
Transport measurements by Minnesota group were supported by the U.S. Department of Energy, Office of Science, Basic Energy Sciences, under Award No. ER 46640-SC0002567. 
Device fabrication and characterization by Minnesota group were supported by the NSF Award No. DMR-1309578.
Growth of GaAs/AlGaAs quantum wells at Purdue University was supported by the U.S. Department of Energy, Office of Science, Basic Energy Sciences, under Award No. DE-SC0006671. 
Growth of GaAs/AlGaAs quantum wells at Princeton University was in part by the Gordon and Betty Moore Foundation’s EPiQS Initiative, Grant No. GBMF9615 to L. N. P., and by the National Science Foundation MRSEC Grant No. DMR 1420541. 
A portion of this work was performed at the National High Magnetic Field Laboratory, which is supported by National Science Foundation Cooperative Agreements No. DMR-1157490 and No. DMR-1644779 and the State of Florida.
\end{acknowledgments}

\small{$^\S$X.F. and Q.S. contributed equally to this work.}


\end{document}